# Efficient route to achieve superconductivity improvement via substitutional La-Ce alloy superhydride at high pressure

Jingkai Bi[1], Yuki Nakamoto[2], Katsuya Shimizu[2], Mi Zhou[1], Hongbo Wang[1], Guangtao Liu[1], and Yanming Ma[1, 3]

[1]*State Key Laboratory of Superhard Materials and International Center of Computational Method & Software, College of Physics, Jilin University, Changchun 130012, China*
[2]*Center for Quantum Science and Technology under Extreme Conditions, Osaka University, Toyonaka, Osaka 560-8531, Japan*
[3]*International Center of Future Science, Jilin University, Changchun 130012, China*

The discovery of clathrate superhydrides has approached the long-standing dream of room-temperature superconductivity and thus inspired their prosperous research under high pressure. However, how to experimentally optimize these compelling superhydrides is still a formidable challenge. Here, we find that half of the Ce atoms in the recently discovered hexagonal close packed (*hcp*) $CeH_9$ structure can be randomly replaced by adjacent La, resulting in the formation of $LaH_9$ unit that is impossible in a binary system. Our experiments show that *hcp* $(La, Ce)H_9$ can be synthesized at ~110 GPa and possesses a maximum $T_c$ of 178 K at higher pressure, which is evidenced by in-situ X-ray diffraction and electronic transport measurement where a sharp drop of resistivity to zero and a characteristic decrease of $T_c$ under a magnetic field up to 9 T. More importantly, the $T_c$ of $(La, Ce)H_9$ is significantly increased by ~50-80 K compared to $CeH_9$, showing the hitherto highest $T_c$ at megabar pressure. Our experimental results not only verify the feasibility of improving the superconductivity of hydrides by introducing other suitable metals, but also provide important inspiration for finding high-$T_c$ superconductors in various multinary superhydrides.

Email: liuguangtao@jlu.edu.cn

**Introduction**

Since the first discovery of mercury superconductivity, the search for room-temperature superconductors has been one of the most important topics in the field of condensed matter physics [1-3]. As the most promising candidate for high-temperature superconductors, compressed metallic hydrogen-rich compounds have achieved successive breakthroughs in superconducting temperature ($T_c$) records: 203 K of covalent $SH_3$ [4-6], 250-260 K of ionic $LaH_{10}$ [7-10], and possible 288 K of the vague C-S-H system [11] that has caused widespread controversy [12-14] and needs further experiment to clarify its stoichiometry and crystal structure.

Since the clathrate superhydride $CaH_6$ was first theoretically proposed in 2012 [15], the great success of the clathrate family represented by $CaH_6$ [16], $YH_6$ [17], $YH_9$ [18,19], $CeH_9$, $CeH_{10}$ [20-22], and $LaH_{10}$ [7-10] *et al.* has sparked a research boom in past years owing to their vast array of compound types with diverse central metals. Overall, a great number of theoretical and experimental works have indicated that the atomic radius, electronegativity, and valence electron of metal play a pivotal role in tuning the concerned properties of superhydrides (*e.g.*: superconductivity and stability) [23]. Among these superhydrides, $LaH_{10}$ possesses the highest $T_c$ of 250-260 K but its synthesis pressure is harsh (> 150 GPa); in contrast, $CeH_9$ or $CeH_{10}$ with relatively low $T_c$ of 95-115 K can be obtained around megabar pressure. How to tune and improve the superconducting properties of these superhydrides, while taking into account their synthesis pressures, is one of the most important scientific issues in the field.

With the purpose of ameliorating superconducting hydrides, the scientists thus shifted their research focus to ternary systems that have a higher degree of freedom and include more candidate structural prototypes for superconductivity screening. Since doped Li donates an extra electron to molecular-like hydrogen in $MgH_{16}$, hydrogen in $Li_2MgH_{16}$ exhibits atomic-like characteristics, and its $T_c$ is estimated to be above 400 K at 250 GPa, demonstrating the potential for superconductivity beyond room temperature in ternary superhydrides [24]. On the other hand, sodalite-like $XYH_8$ structures, containing metals of differing sizes, show a remarkable advantage in dynamical stability that may maintain the superconducting phase of ternary hydrides to relatively low pressures [25-27]. Recent theoretical studies have suggested more ternary hydrides, such as Ca-Y-H [28], Ca-Mg-H [29], that are expected to possess desirable physical features and prospects.

Since rare-earth metals have similar electronegativities, electron configurations, and atomic radii, their disordered solid solution alloys are easy to form [30,31]. Reasonably, part of the metal atoms of binary superhydrides can be randomly substituted with other similar rare earth metals, forming a ternary alloy superhydride that shares the same crystal structure as their binary ones. Previous experimental work has proved that $LaH_6$ and $YH_{10}$ units, which are unreachable in binary systems, do appear in La-Y alloy hydrides at pressures of 170–196 GPa [32]. Therefore, it is of great interest to explore the combination of the two most representative cubic La-H [9,10,33] and hexagonal close packed (*hcp*) Ce-H [20-22] systems in rare earth superhydrides, and check whether the anticipated ternary La-Ce alloy superhydride with high $T_c$ can be realized in an experiment.

In this work, we experimentally investigated the crystal structure, superconductivity, and pressure stability range of La-Ce alloy superhydride. Substitutional *hcp* (La, Ce)$H_9$ (occupancy: ~0.5 for La and ~0.5 for Ce) was successfully synthesized with the starting materials of equiatomic La-Ce alloy and ammonia borane ($NH_3BH_3$) at about 110 GPa and 2,100 K and could be maintained to at least 90 GPa during decompression. Compared with binary $CeH_9$, the $T_c$ is dramatically increased by 50-80 K in ternary (La, Ce)$H_9$, where an unprecedented $LaH_9$ unit is stabilized and makes a major contribution to its high-temperature superconductivity. The present results reveal that multinary alloy superhydride is a promising superconductive system that may usher in another breakthrough in the clathrate superhydrides, through a careful choice of the metal elements combination.

**Results and Discussion**

We took a flake of alloy sample from a homogeneous region of the inside part of the specimen and loaded it into a diamond anvil cell (DAC) together with $NH_3BH_3$ as the pressure transport medium and hydrogen source. The desired mole ratio of La: Ce was determined to be ~1:1 by energy dispersive X-ray (EDX) spectroscopy (Fig. S1). To measure the synchrotron X-ray diffraction (XRD) and $T_c$ of the synthesized La-Ce superhydride, a total of 7 samples were compressed to 110-120 GPa at room temperature and then heated to 2,100 K with pulsed radiation from an yttrium-aluminum-garnet laser. The color of the sample changed significantly after laser spot irradiation (Fig. 2a), indicating that the expected chemical reaction occurred. After keeping the synthesized samples under high pressure and quenching to room temperature, we performed subsequent structure and superconductivity characterizations.

To determine the crystal structure of synthesized superhydride, we conducted in-situ XRD experiments in synchrotron radiation sources. The representative XRD pattern of the product in cell-4 at 110 GPa is shown in Fig. 1. We found that these observed peaks can be indexed by an *hcp* lattice with cell parameters of $a$ = 3.76 Å and $c$ =5.68 Å. The weak peak marked with an asterisk is from the tetragonal tetrahydride reported in previous studies [20,21], which is a common occurrence caused by temperature or pressure gradients during high-temperature-temperature synthesis. A similar *hcp* structure mixed with tetrahydride was successfully reproduced at 115 GPa in another independent experimental run (Fig. S2), corroborating the reliability of this structure. Although the occupancy details of the hydrogen atoms cannot be determined experimentally due to the weak X-ray scattering cross section, the measured unit cell volume can be used to estimate the stoichiometry of the superhydride, and the ratio of H/metal was estimated to be ~ 9 through the lattice volume expansion from 15.95/14.07 Å$^3$/f.u. (elemental La/Ce lattice) to 34.80 Å$^3$/f.u.. Furthermore, as shown in Fig. S3, the fitted equation of state (EOS) during decompression is highly consistent with the simulated EOSs, where the experimental points lie between the lines of *hcp* CeH$_9$ and hypothetical *hcp* LaH$_9$. Combined with the composition of the original alloy sample, our diffraction experiments show that ~50% of the Ce atoms in the recently discovered structure of $P6_3/mmc$-CeH$_9$ can be replaced by La, forming a unique ternary alloy superhydrides $P6_3/mmc$-(La, Ce)H$_9$.

We synthesized this unique ternary alloy superhydride at 110 GPa, encouraging us to further perform its electrical transport measurements at low temperatures. Representative electrical resistance measurements as a function of temperature at high pressures are shown in Fig. 2b, which clearly shows the superconducting transition. Superconducting transitions with apparently sharp drops in resistance occur at 158, 168, and 173 K at about 110, 125, and 110 GPa, respectively. In these experiments, completely zero resistance states were observed in cell-1 and cell-4 (Fig. 2b inset), excluding the possibility that the abrupt drop of resistance on cooling arises from structural transitions. To determine the highest value of $T_c$, we proceeded to regulate the pressure dependence on $T_c$, as shown in Fig. 3. With the increase of pressure, $T_c$ increases first and then tends to be flat. And the highest $T_c$ of 178 K in this work was observed at 172 GPa. In addition, the pressure dependence of $T_c$ varies slightly in different experiments, which may originate from subtle differences in the molar ratios of the La-Ce alloy samples, or different degrees of anisotropic stress during compression/decompression that leads to variable deformation of the lattice in different experiments [34]. With decreasing pressure, the superconducting transition tends to disappear at about 90 GPa (Fig. S6), indicating a probable decomposition of the superconducting phase.

Due to the rather small size of the samples, it is almost impossible to detect the weak signal of the magnetic flux expulsion effect with current experimental capabilities [10,19]. However, due to the Pauli paramagnetic effect of electron spin polarization and the diamagnetic effect of orbital motion, the applied external field can disrupt the Cooper pairs, thereby reducing the value of $T_c$. As shown in Fig. 4, the resistance drop gradually shifts to lower temperatures as the magnetic field increases in the 0–9 T range at 110 GPa. The upper critical field as a function of temperature, defined as 90% of the resistance, is shown in the inset of Fig. 4b. At $\mu_0H$ = 9 T, the application of the magnetic field reduces $T_c$ by about 12 K. The extrapolated values of the upper critical field $\mu_0H_{c2}(T)$ and the coherence lengths towards $T = 0$ K are 56 T and 24 Å and 76 T and 21 Å, respectively, by Ginzburg-Landau (GL) [35] and Werthamer-Helfand-Hohenberg (WHH) [36] model fits. The magnitude of the short coherence lengths and high upper critical fields indicate ternary (La, Ce)H$_9$ is a typical type-II superconductor.

Since the clathrate monohydride XH$_9$ phase with H$_{29}$ cage was predicted to be a potential high-temperature superconductor in 2017 [7,8], this structural prototype was experimentally verified by a list of rare earth superhydrides in succession. In particular, *hcp* CeH$_9$ was synthesized at easily achievable megabar pressure and exhibited an experimental $T_c$ of ~95 K [20], basically coinciding with the estimated values (56-75 K) of theoretical works [8, 21]. Compared to high-$T_c$ cubic LaH$_{10}$ [9,10,33], the low $T_c$ of CeH$_9$ originates from its low logarithmic average phonon frequency, which should be related to the localization of 4$f$ electron of Ce under high pressure. Furthermore, no superconductivity or only low-temperature superconductivity was observed in heavy rare earth superhydrides, where the strongly correlated interactions of 4$f$ electrons suppress the superconductivity. Consequently, it is reasonable that the superconductivity is enhanced by replacing Ce of CeH$_9$ with La that has no 4$f$ electron.

It is also worth mentioning that the LaH$_9$ unit appears in this ternary hydride even if binary *hcp* LaH$_9$ is not thermodynamically stable at the corresponding pressure (~100 GPa), as happened in the La-Y-H system [32]. In addition to crossing the energy barrier, laser heating is crucial in stabilizing substitutional (La, Ce)H$_9$, in which the contribution of configuration entropy to the total free energy would overcome the contribution of mixing enthalpy and the energy increment of LaH$_9$ relative to LaH$_{10}$ at a sufficiently high temperature [31]. Especially, the fact that the configuration entropy reaches the maximum in an equal-atomic system suggests that the La-Ce alloy used is beneficial to the synthesis in this study.

**Summary**

In conclusion, a high-$T_c$ (La, Ce)H$_9$ was experimentally discovered at pressures above 110 GPa, and exhibits $T_c$ of 143-178 K in the pressure range of the current study. Compared with binary systems, this ternary superhydride integrates the respective advantages of La-H and Ce-H, thus showing remarkable characteristics in both stability and superconductivity. These results suggest that substitutional alloy superhydrides are promising superconductors, taking an effective step towards the search for high-temperature superconductivity at relatively lower pressure. The present findings also expand the scope of ongoing studies in search of room-temperature superconductors among more multinary systems.

**Experimental Details**

La-Ce alloy (1:1 in molar ratio) was prepared by levitation melting technology, where mixed elemental La and Ce with purities of 99.7% were heated to completely melt in an Ar atmosphere at a pressure of 60 KPa for 1-2 minutes and were annealed for about 40 minutes. To ensure the uniformity of the composition, the specimen ingot needed to be turned over and the above process was repeated more than 2 times.

For sample synthesis, we synthesized hydride via a reaction of La-Ce alloy and NH$_3$BH$_3$ (Sigma-Aldrich, 97%) in diamond anvil cells (DACs). The diamond anvils with culets of 60-100 $\mu$m in diameter were beveled at 8.5° to about 250 $\mu$m in diameter. A composite gasket consisting of a rhenium outer annulus and an aluminum oxide (Al$_2$O$_3$) epoxy mixture insert was employed to contain the sample while isolating the electrical leads in the electrical measurements. La-Ce alloy foil with a thickness of 2-3 $\mu$m was sandwiched between the NH$_3$BH$_3$. Sample preparation and loading were done in an inert Ar atmosphere with residual O$_2$ and H$_2$O contents of < 0.01 ppm to guarantee that the sample was properly isolated from the surrounding air atmosphere. The samples were compressed to target pressures at room temperature. The pressure values in cells were determined from room-temperature first-order diamond Raman edge calibrated by Akahama [37,38]. One-side laser-heating experiments were performed using a pulsed YAG laser (1064 nm) with spots of approximately 10 $\mu$m in diameter. The temperature was determined using the emission spectrum of the black body radiation within Planck's radiation law.

X-ray diffraction (XRD) patterns were obtained at BL15U1 ($\lambda$ = 0.6199 Å) of Shanghai Synchrotron Radiation Facility and BL10XU ($\lambda$ = 0.4131 Å) of at the SPring-8 facility [39] with focused monochromatic X-ray beams (5 × 12 and 3 × 2 $\mu m^2$). A Mar165 CCD detector and an imaging plate detector (RAXIS-IV; Rigaku) were used to collect the angle-dispersive XRD data, respectively. The sample to detector distance and other geometric parameters were calibrated using a $CeO_2$ standard. The software package Dioptas was used to integrate powder diffraction rings and convert the 2-dimensional data to 1-dimensional profiles [40]. The full profile analysis of the diffraction patterns and the Rietveld refinements were done using GSAS and EXPGUI packages [41].

The resistance was all measured via the four-probe van der Pauw method where four Au electrodes were placed on the $NH_3BH_3$ plate with currents of 1-100 $\mu$A. The temperature dependence of the electrical resistance was measured upon cooling and warming cycles with a slow temperature ratio (0.2 K·min$^{-1}$). The data was taken upon warming, as it yielded a more accurate temperature reading. Symmetric Cu-Be alloy DACs were used for electronic transport measurements under external magnetic fields up to 9 T.

**Acknowledgements**

This research was supported by National Key Research and Development Program of China (Grant No. 2021YFA1400203), the National Natural Science Foundation of China under Grants No. 12074139, 12074138, and 52090024; Interdisciplinary Integration and Innovation Project of Jilin University (JLUXKJC2020311), Fundamental Research Funds for the Central Universities (Jilin University, JLU), Program for JLU Science and Technology Innovative Research Team, the Strategic Priority Research Program of Chinese Academy of Sciences (Grant No.

XDB33000000). The work was also supported by JSPS KAKENHI Grant Numbers 20H05644. The XRD measurements were performed at the BL15U1 at Shanghai Synchrotron Radiation Facility and the BL10XU of SPring-8 (Proposal No. 2021A1172 and 2021B1407).

**Figures and captions:**

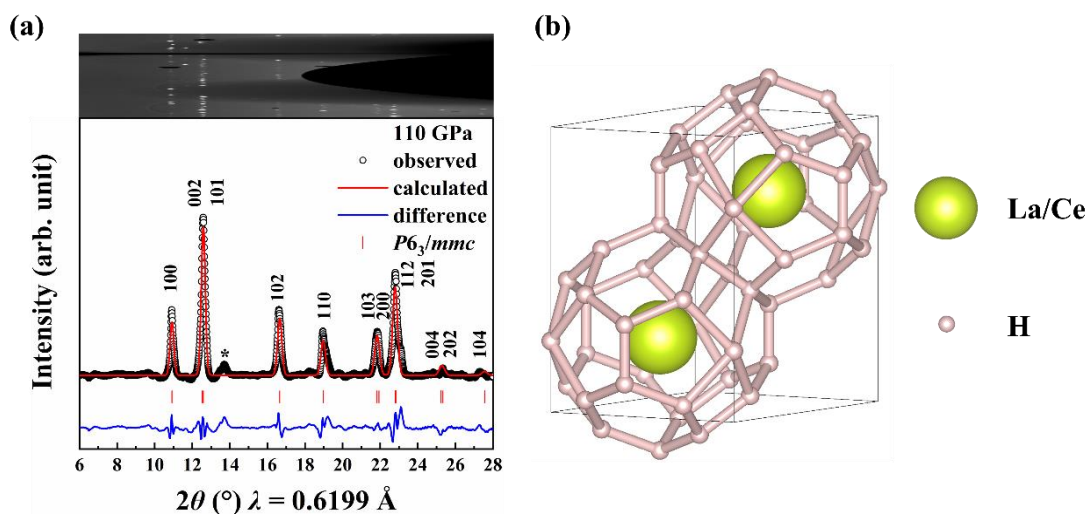

**FIG. 1 Structural data for (La, Ce)H₉ synthesized from La-Ce alloy and NH₃BH₃.** (a) Synchrotron X-ray diffraction pattern of the La-Ce alloy hydrides obtained following laser heating and the Rietveld refinement of the $P6_3/mmc$-(La, Ce)H₉ structure at 110 GPa. (b) Crystal structure of $P6_3/mmc$-(La, Ce)H₉.

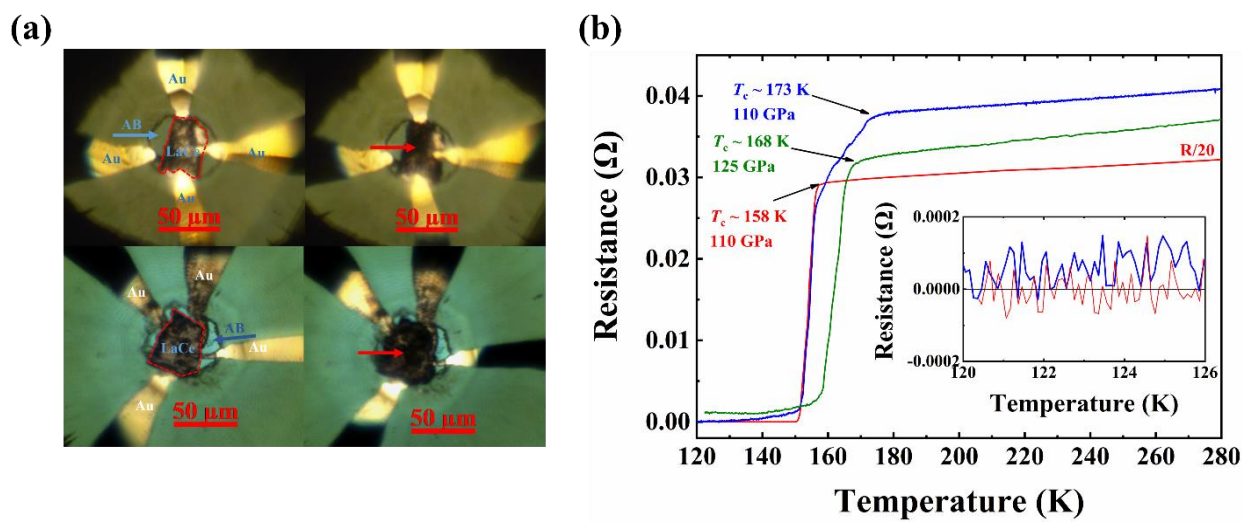

**FIG. 2 Superconducting transitions determined by the electrical resistivity measurements in typical cells. (a)** Pressurized La-Ce alloy with NH₃BH₃ (AB) and four electrodes on the insulated gasket with double-side illumination. The edge of La-Ce alloy is marked with red dotted lines and the red arrows point to the parts with apparent changes after heating. **(b)** Resistance measurements of synthesized (La, Ce)H₉.

Red curve: sample at 110 GPa with $T_c \sim 158$ K. Green curve: sample at 125 GPa with $T_c \sim 168$ K. Blue curve: sample at 110 GPa with $T_c \sim 173$ K. The resistance near zero is shown on a smaller scale in the left bottom inset.

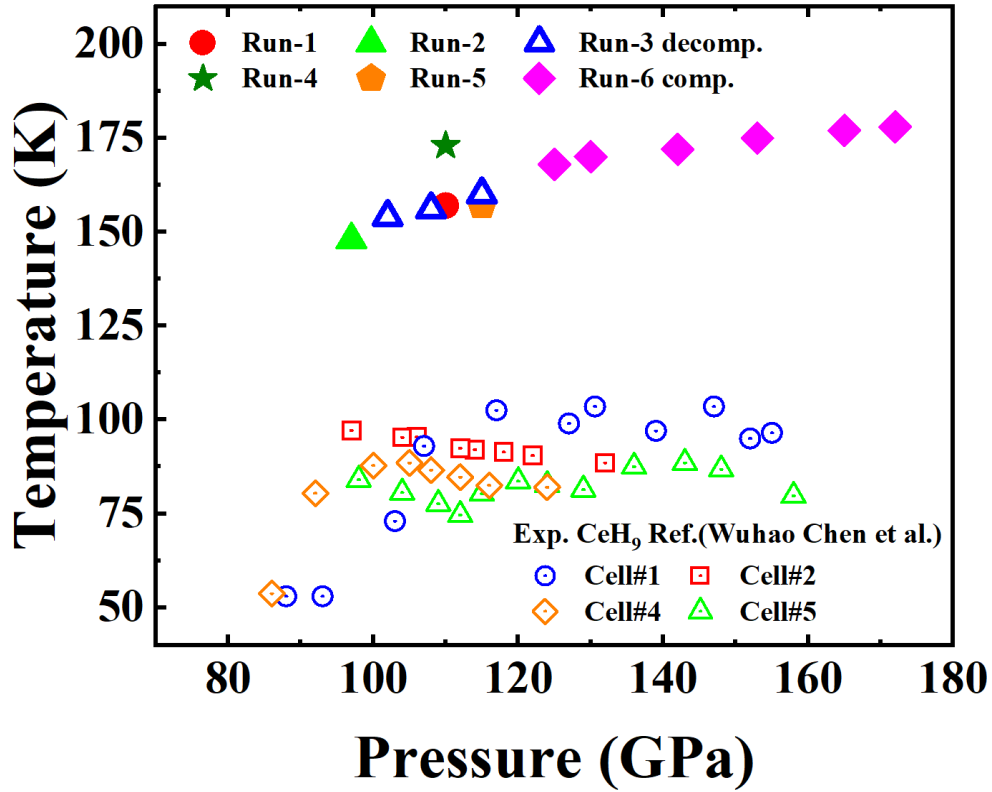

**FIG. 3 $P$- $T_c$ date of $P6_3/mmc$-(La, Ce)H$_9$.** The dependence of the critical temperature $T_c$ on pressure; the results from six different experiments are marked in different colors.

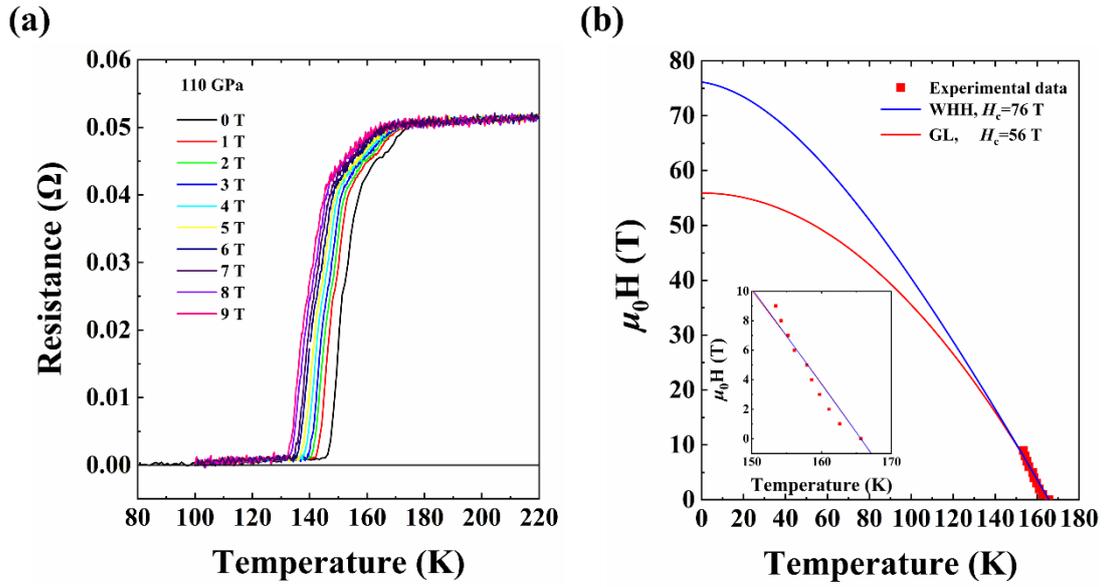

**FIG. 4 Electrical measurements in the external magnetic field of the synthesized (La, Ce)H₉.** (a) Temperature dependence of the electrical resistance under applied magnetic fields of H = 0-9 T at 110 GPa. (b) Upper critical field $H_{c2}$ versus temperature following the criterion of 90% of the resistance in the metallic state at 110 GPa, fitted with the GL and WHH models. Inset: the dependence of the $T_c$ under the applied magnetic field.

Supplementary Material

# Efficient route to achieve superconductivity improvement via substitutional La-Ce alloy superhydride at high pressure

J Jingkai Bi[1], Yuki Nakamoto[2], Katsuya Shimizu[2], Mi Zhou[1], Hongbo Wang[1], Guangtao Liu[1], and Yanming Ma[1, 3]

[1]*State Key Laboratory of Superhard Materials and International Center of Computational Method & Software, College of Physics, Jilin University, Changchun 130012, China*
[2]*Center for Quantum Science and Technology under Extreme Conditions, Osaka University, Toyonaka, Osaka 560-8531, Japan*
[3]*International Center of Future Science, Jilin University, Changchun 130012, China*

Table. S1 Detailed parameters of the DACs used in the experiments.

| Number | Culet size (μm) | Gasket | Composition | Temperature (K) | Pressures (GPa) | Measurement |
|---|---|---|---|---|---|---|
| #Cell-1 | 80 | Re+Al$_2$O$_3$/epoxy | La-Ce alloy+ NH$_3$BH$_3$ | 2100 | 110 | SC*, Run-1, XRD |
| #Cell-2 | 80 | Re+Al$_2$O$_3$/epoxy | La-Ce alloy+ NH$_3$BH$_3$ | 2200 | 97 | SC, Run-2 |
| #Cell-3 | 80 | Re+Al$_2$O$_3$/epoxy | La-Ce alloy+ NH$_3$BH$_3$ | 2100 | 85-115 | SC, Run-3 |
| #Cell-4 | 80 | Re+Al$_2$O$_3$/epoxy | La-Ce alloy+ NH$_3$BH$_3$ | 2300 | 110 | SC, Run-4, XRD, MF** |
| #Cell-5 | 80 | Re+Al$_2$O$_3$/epoxy | La-Ce alloy+ NH$_3$BH$_3$ | 2200 | 115 | SC, Run-5 |
| #Cell-6 | 60 | Re+Al$_2$O$_3$/epoxy | La-Ce alloy+ NH$_3$BH$_3$ | 2100 | 125-172 | SC, Run-6 |
| #Cell-S1 | 80 | Re+Al$_2$O$_3$/epoxy | La-Ce alloy+ NH$_3$BH$_3$ | 2100 | 115 | XRD |

* Superconductivity electrical transport measurement (SC).

** Transport measurements under varying external magnetic fields (MF).

Table. S2 Crystal structure information of (La, Ce)H$_9$.

| Number | P (GPa) | Structure | Lattice parameter (Å) | Unit cell volume (Å$^3$) |
|---|---|---|---|---|
| #Cell-4 | 110 (experiment) | $P6_3/mmc$ (La, Ce)H$_9$ | $a$ = 3.76 $c$ = 5.68 | 34.80 |

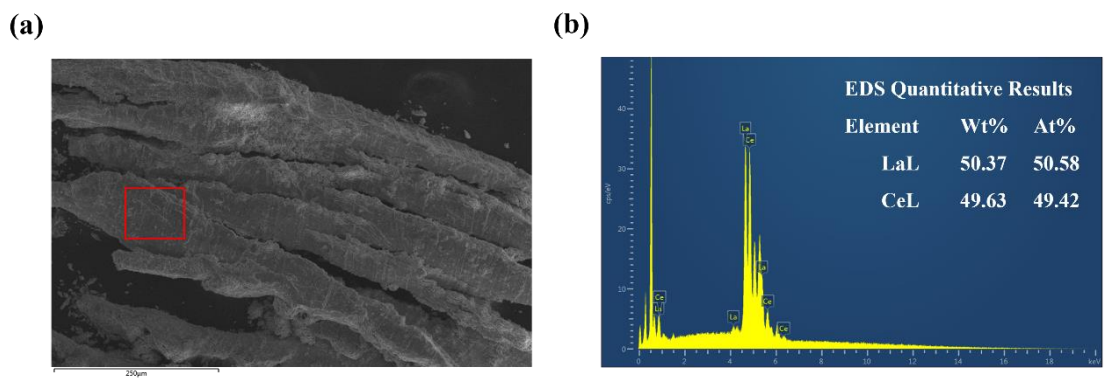

**FIG. S1 La-Ce alloy scanning electron microscopy image and energy-dispersive X-ray spectrum.** Representative (a) scanning electron microscope (SEM) image of La-Ce alloy at 0 GPa and corresponding (b) energy-dispersive X-ray (EDX) spectrum from the area highlighted by the red rectangle in the SEM image.

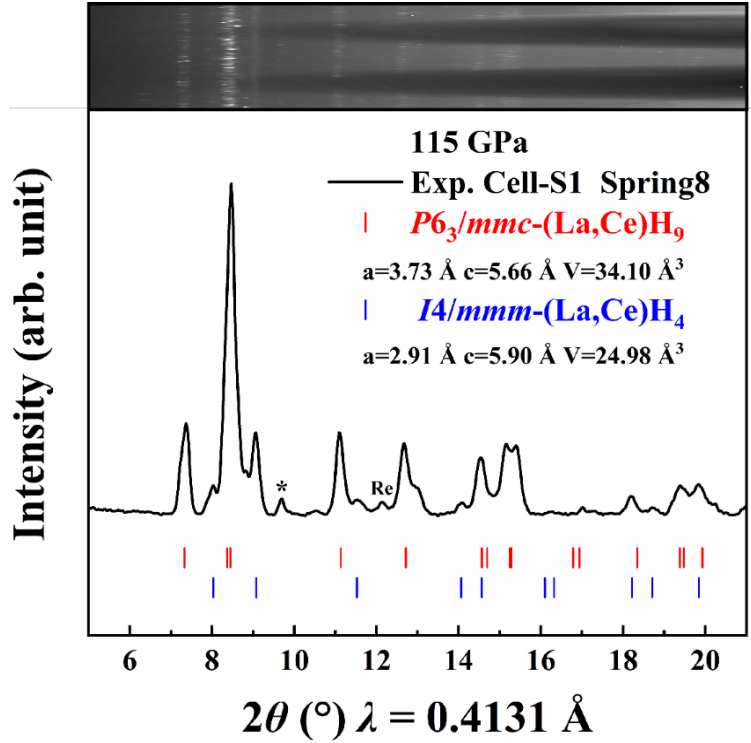

**FIG. S2** Phase analysis of the composition in Cell-S1 according to synchrotron X-ray diffraction (0.4131 Å) data. The weak peak marked with an asterisk may be from the undetermined hydride impurity.

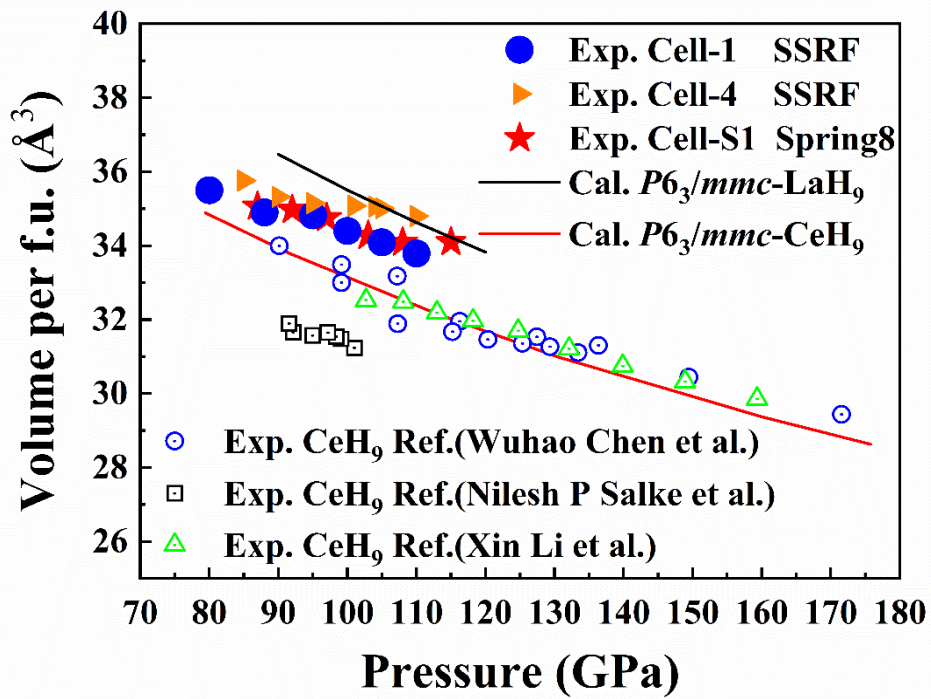

**FIG. S3** Experimental equation of state (EOS) for different samples and calculated curves of LaH$_9$, YH$_9$, and (La, Ce)H$_9$.

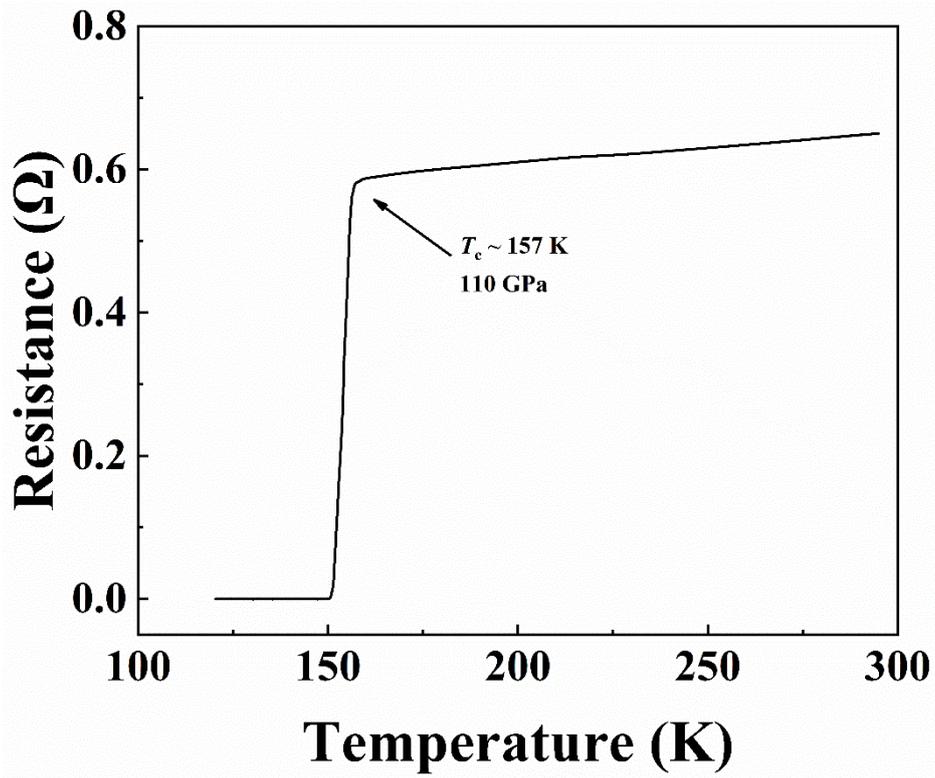

**FIG. S4** Superconducting transitions of warming cycle in Run-1.

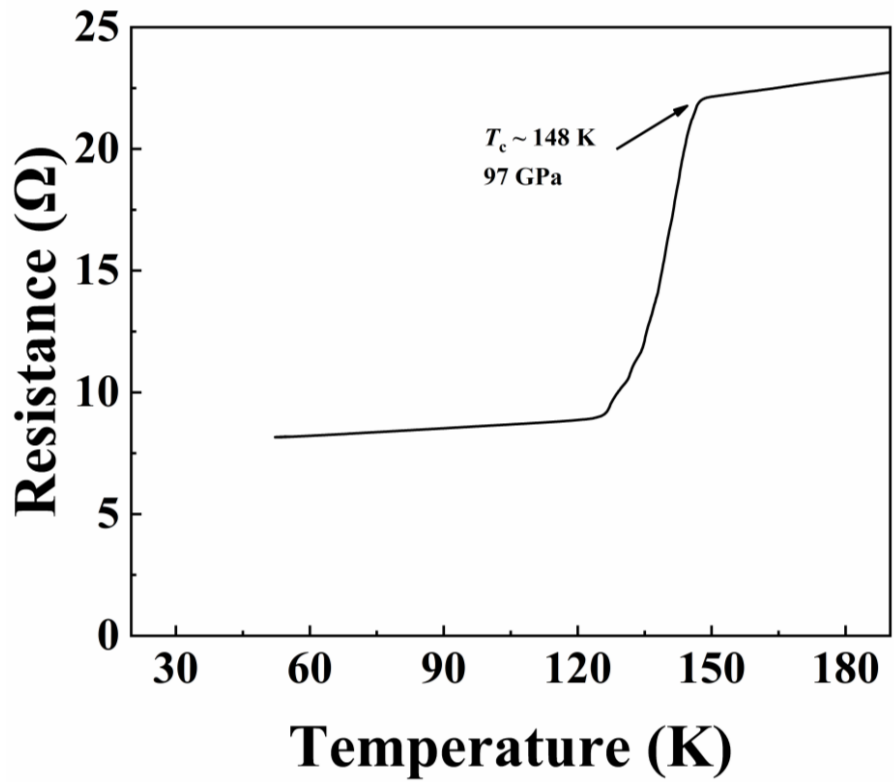

**FIG. S5** Superconducting transitions of warming cycle in Run-2.

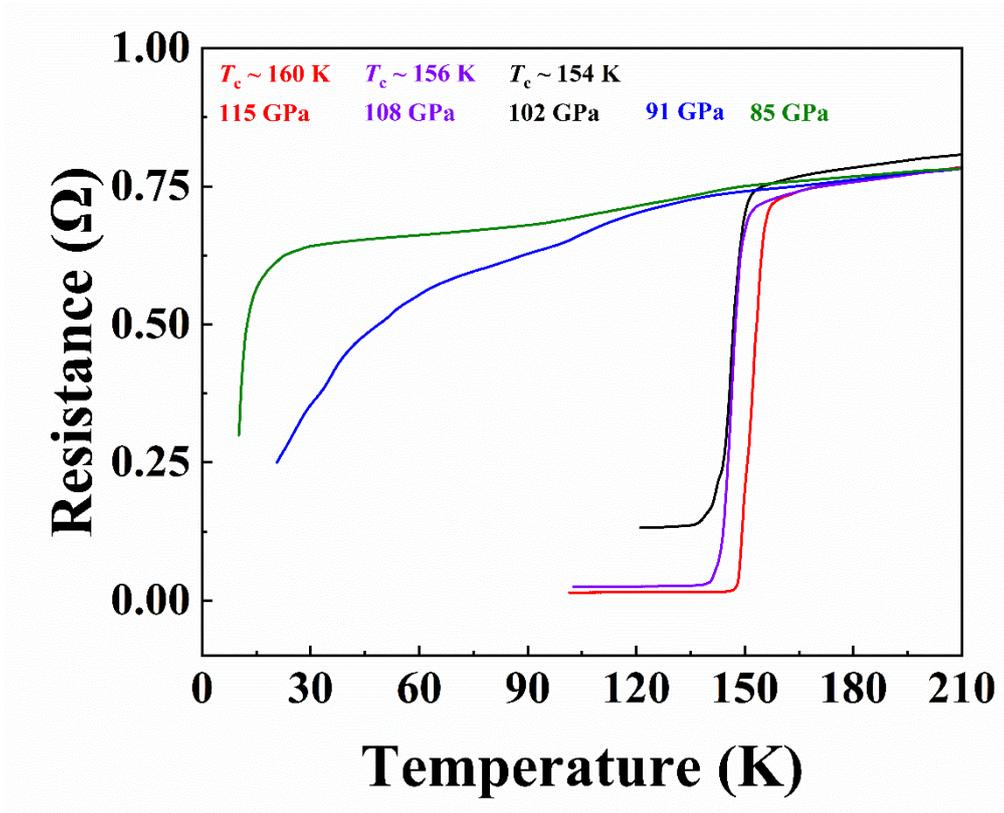

**FIG. S6** Superconducting transitions of warming cycles in Run-3.

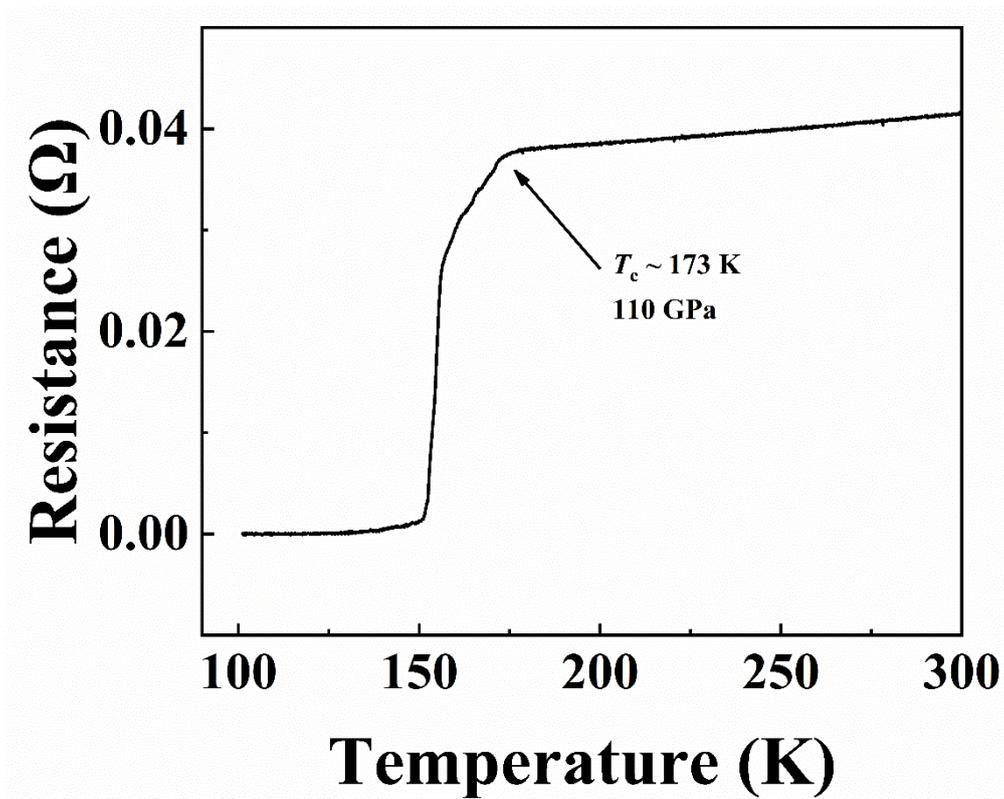

**FIG. S7** Superconducting transitions of warming cycle in Run-4.

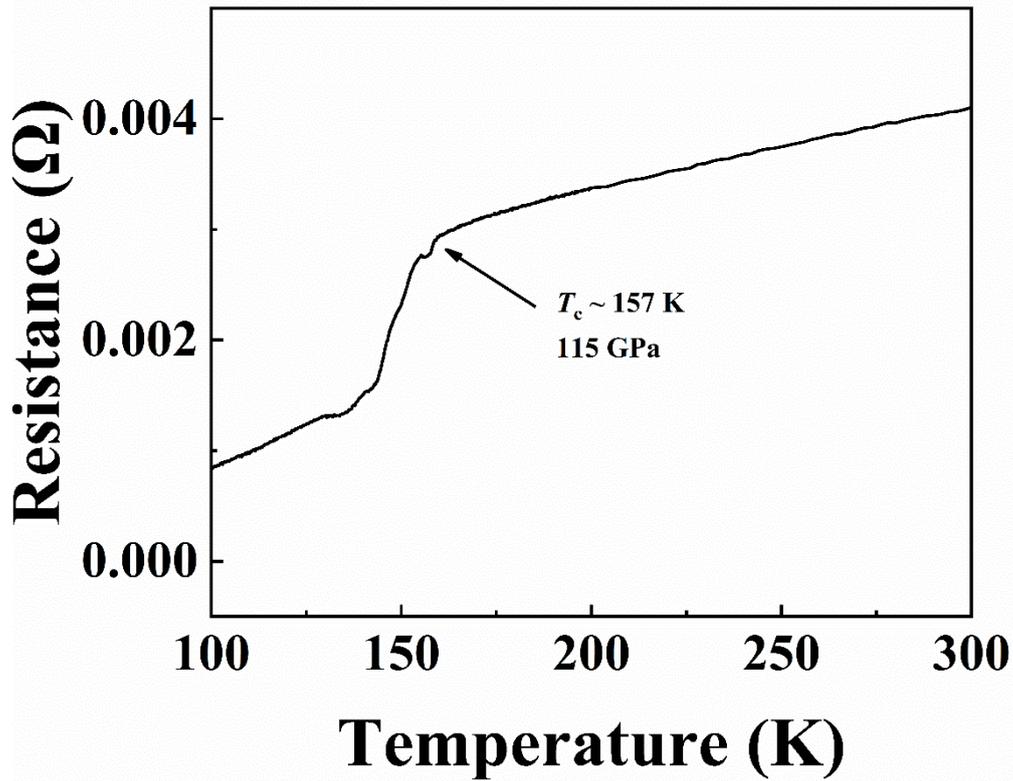

**FIG. S8** Superconducting transitions of warming cycle in Run-5.

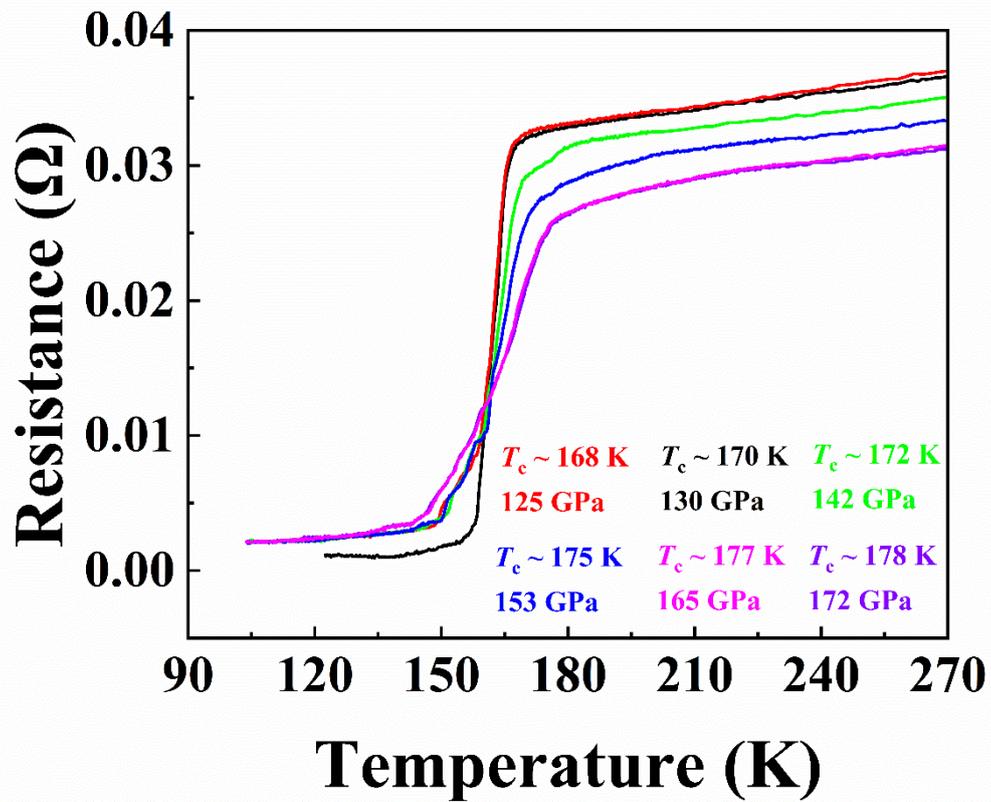

**FIG. S9** Superconducting transitions of warming cycles in Run-6.